\documentclass[twocolumn,superscriptaddress,floatfix,showpacs,prl,longbibliography,10pt]{revtex4-1}
\usepackage[english]{babel}
\usepackage{graphicx}
\usepackage{amsmath}
\usepackage{amssymb}
\usepackage{bm}
\usepackage{color}

\begin{document}

\title{Interplay of exciton condensation and quantum spin Hall effect in InAs/GaSb~bilayers}
\author{D. I. Pikulin}
\affiliation{Instituut-Lorentz, Universiteit Leiden, P.O. Box 9506, 2300 RA Leiden, The Netherlands}
\author{T. Hyart}
\affiliation{Instituut-Lorentz, Universiteit Leiden, P.O. Box 9506, 2300 RA Leiden, The Netherlands}


\begin{abstract} 
We study the phase diagram of the inverted InAs/GaSb bilayer quantum wells. For small tunneling amplitude between the layers, we find that the system is prone to formation of an $s$-wave exciton condensate phase, where the spin-structure of the order parameter is uniquely determined by the small spin-orbit coupling arising from the bulk inversion asymmetry. The phase is topologically trivial and does not support edge transport. On the contrary, for large tunneling amplitude, we obtain a topologically non-trivial quantum spin Hall  insulator phase with a $p$-wave exciton order parameter, which enhances the hybridization gap. These topologically distinct insulators are separated by an insulating phase with spontaneously broken time-reversal symmetry. Close to the phase transition between the quantum spin Hall and time-reversal broken phases, the edge transport shows quantized conductance in small samples, whereas in long samples the mean free path associated with the backscattering at the edge is temperature independent, in agreement with recent experiments.
\end{abstract}

\maketitle

\textit{Introduction.--}  Two-dimensional quantum spin Hall (QSH) insulators  are topologically distinguishable from  conventional insulators due to a non-trivial topological invariant arising from band inversion \cite{Kane05, Zhang06, Liu08}. The conducting and valence bands in QSH insulators are connected by gapless helical edge modes, which are protected against elastic backscattering from time-reversal symmetric perturbations. Recent experimental advances have revealed two materials, HgTe/CdTe \cite{Konig07, Roth09, Konig13} and InAs/GaSb \cite{Du11, Du13} quantum wells, where the existence of helical edge states have been confirmed. In addition to the unique electrical properties arising due to edge modes \cite{Kane05, Zhang06, Konig07, Roth09, Konig13, Du11, Du13}, these materials in proximity to superconductors are interesting as a  platform for Majorana zero-modes \cite{Fu08, Fu09} and flux-controlled quantum information processing \cite{Mi13,Braiding}.

The recent observation of the QSH effect in InAs/GaSb bilayers \cite{Du13} is theoretically puzzling, because  conductance quantization was found  up to magnetic field on the order of 10 T in short samples. On the other hand, even in the absence of magnetic field the longitudinal resistance in long samples increased linearly with the device length. While inelastic processes can in principle give rise to a finite mean free path associated with the backscattering at the edge,  the existing theoretical models  \cite{Maciejko09, Tanaka11, Budich12, Crepin12, Lezmy12, Ilan12, Schmid12, Vayrynen13} do not explain the observation that the mean free path  was found to be temperature independent at least for a temperature range  20 mK - 4.2 K \cite{Du13, Knez13, Nowack14}. Therefore, we expect that  non-perturbative effects due to disorder or interactions beyond the existing approaches are important. In particular, the temperature-independent mean free path indicates that the dominating backscattering process might be an elastic one, which is allowed if time-reversal symmetry is either dynamically or spontaneously broken. While mechanisms resulting in dynamical time-reversal symmetry breaking have been proposed \cite{Lunde12, Maestro13}, it is unlikely that they could account for the experimentally observed mean free path $\sim 4 \ \mu$m.  

In this Letter, we consider the influence of exciton condensation on the QSH effect in InAs/GaSb bilayers. While exciton condensation is theoretically predicted in inverted type II electron-hole bilayers \cite{Keldysh, Lozovik, Lozovik1, Zhu95, Naveh96, Littlewood, MacDonald09}, such as InAs/GaSb quantum wells, an unambiguous observation of a thermodynamically stable exciton condensate phase in these systems is a long-standing problem. Indeed, so far the best studied  exciton condensate phase is the quantum Hall bilayer state at half-filled  Landau levels \cite{Fertig89, Murphy94}, where the ability to separately contact the two layers has allowed to probe the order parameter in terms of counterflow superfluidity along the layers and Josephson-like tunneling between the layers  \cite{Spielman00, Spielman01, Yoon, Corbino-Eisenstein, Huang12}. Here, we go beyond the earlier theoretical models for exciton condensates \cite{Keldysh, Zhu95, Naveh96, Littlewood, MacDonald09} by studying the spin-structure of the order parameter, when the relevant spin-orbit and tunneling terms for the InAs/GaSb bilayers are taken into account. For small tunneling amplitude we find a topologically trivial $s$-wave exciton condensate phase, whereas for relatively large tunneling we obtain a topologically non-trivial QSH insulator phase. These topologically distinct insulators are separated by an insulating phase with spontaneously broken time-reversal symmetry. Close to the phase transition between the QSH and time-reversal broken phases, the conductance is quantized in small samples, whereas the mean free path in long samples is  temperature-independent for a wide range of temperature, in agreement with the recent experiments \cite{Du13}.

\textit{Model.--}  We consider a bilayer electron-hole systems described by Hamiltonian $\hat{H}=\hat{H}_0+\hat{H}_I$, where $\hat{H}_0$ is single particle Hamiltonian for the InAs/GaSb quantum wells (see below) and the Coulomb interaction between the electrons is described by the Hamiltonian
\begin{equation}
\hat{H}_I= \frac{1}{2} \sum_{a, a', s, s'} \sum_{\mathbf{k}, \mathbf{k}', \mathbf{q}} V^{aa'}(q) c_{\mathbf{k}sa}^\dag c_{\mathbf{k}'s'a'}^\dag c_{\mathbf{k}'+\mathbf{q}s'a'} c_{\mathbf{k}-\mathbf{q}sa},
\end{equation}
where $V^{aa}(q)=e^2 F^{aa}(q)/(2 \epsilon \epsilon_0 L^2 q)$, $V^{12}(q)=V^{21}(q)=e^2  F^{12}(q) e^{- qd}/(2 \epsilon \epsilon_0 L^2 q)$
and $F^{ab}$ are the structure factors for the layers of thicknesses $W_{1,2}$, see \cite{Naveh96}. 

By performing a mean field approximation for the interaction term $\hat{H}_I$ and including the terms of the single-particle Hamiltonian $\hat{H}_0$  \cite{Liu08}, we arrive to a mean-field  Hamiltonian
\begin{equation}
\hat{H}_{\rm mf}= \sum_{\mathbf{k}} \psi_{\mathbf{k}}^\dag H (\mathbf{k}) \psi_{\mathbf{k}}, \  
H(\mathbf{k})=\begin{pmatrix}
H_{11}(\mathbf{k}) & H_{12}(\mathbf{k}) \\
H_{12}^\dag(\mathbf{k}) & H_{22}(\mathbf{k})
\end{pmatrix},
 \end{equation}
where the Hamiltonian for each layer is given by
\begin{eqnarray}
H_{11}(\mathbf{k})&=&\bigg[\frac{\hbar^2 k^2}{2m_e}-E_G+\epsilon_1^{\rm mf}(\mathbf{k})-\mu\bigg] \sigma_0- \sum_{i=1}^3 h_{1,i}^{\rm mf}(\mathbf{k}) \sigma_i \nonumber\\&&+\Delta_e (k_x \sigma_1 - k_y \sigma_2)+\xi_e(k_y \sigma_1-k_x \sigma_2), \nonumber \\
H_{22}(\mathbf{k})&=&\bigg[E_G-\frac{\hbar^2 k^2}{2m_h}+\epsilon_2^{\rm mf}(\mathbf{k})-\mu\bigg] \sigma_0- \sum_{i=1}^3 h_{2,i}^{\rm mf}(\mathbf{k}) \sigma_i \nonumber\\&&+\Delta_h (k_x \sigma_1 + k_y \sigma_2)
\end{eqnarray}
and the coupling between the layers is described by
\begin{equation}
H_{12}(\mathbf{k})=A (k_x \sigma_3+i k_y \sigma_0)-i \Delta_{z}\sigma_2-\Delta^{\rm mf}(\mathbf{k}).
\end{equation}
Here $E_G$ is the inverted band gap, $\mu$ is the chemical potential and $m_{e (h)}$ are the effective masses.
Because of the $s$- and $p$-like natures of the conduction and valence bands, respectively, the tunneling term $A(k_x \sigma_3+i k_y \sigma_0)$  must be odd in momentum. 
The spin-orbit couplings $\Delta_e$, $\Delta_h$ and $\Delta_z$ arise due to bulk inversion asymmetry, and $\xi_e$ is the Rashba coupling. The mean field potentials should be solved self-consistently from equations \cite{MacDonald09}
\begin{equation}
\epsilon_a^{\rm mf}(\mathbf k)=-\frac{1}{2}\sum_{s, \mathbf{k'}} V^{aa}(\mathbf{k}-\mathbf{k}') [ \rho^{aa}_{ss}(\mathbf{k}')- \rho^{aa}_{0}(\mathbf{k}')],
\end{equation}
\begin{equation}
\mathbf{h}_a^{\rm mf}(\mathbf k)=\frac{1}{2}\sum_{s, s', \mathbf{k'}} V^{aa}(\mathbf{k}-\mathbf{k}') \rho^{aa}_{ss'} (\mathbf{k}') \vec{\sigma}_{s,s'}
\end{equation}
and
\begin{equation}
\Delta_{s, s'}^{\rm mf}=\sum_{\mathbf{k'}} V^{12}(\mathbf{k}-\mathbf{k}') \rho^{21}_{s's}(\mathbf{k}').
\end{equation}
Here 
$\rho^{aa'}_{ss'}(\mathbf{k})=\langle c_{\mathbf{k}s a}^\dag c_{\mathbf{k}s'a'} \rangle$
is the Hartree-Fock density matrix and $\rho^{aa}_{0}(\mathbf{k}) $ is the density matrix for full valence band in the hole layer and empty conduction band in the electron layer  \cite{MacDonald09}.
The mean field potentials $\epsilon_a^{\rm mf}(\mathbf k)$ describe the renormalization of the band structure, whereas $\mathbf{h}_a^{\rm mf}(\mathbf k)$ can account for spontaneous magnetization and the renormalization of the spin-orbit couplings. For our purposes, the most interesting mean field potentials are
$\Delta^{\rm mf}(\mathbf{k})=\sum_{i=0}^4 \Delta_i^{\rm mf} (\mathbf{k}) \sigma_i$, which describe the full spin structure of the exciton condensate order parameter.

The natural length $d_0$ and energy $E_0$ scales of the problem can be determined from the relation 
$
E_0=(m_e^{-1}+m_h^{-1}) \hbar^2/2 d_0^2=e^2/(4\pi \epsilon \epsilon_0d_0).
$
For InAs/GaSb bilayers typical parameters in the regime of band inversion are expected to be $E_0/k_B \sim 100$ K, $d_0 \sim 10$ nm, $m_e/m_h \sim 1$, $A/(E_0d_0) \sim 0.1$, $\Delta_z/E_0 \sim 0.01$, $\xi_e/(E_0d_0) \sim -0.1$, $d/d_0 \ll 1$, $W_a/d_0 \sim 1$ and $\Delta_e, \Delta_h \sim 0.001$  \cite{Liu-unpublished}.
The parameters $E_G$ and $\mu$ describe the densities in the  layers, and can be controlled with gate voltages. The tunneling terms $A$ and $\Delta_z$ are exponentially sensitive to width and height of an insulating barrier between the layers. 

\textit{Results.--}For $A=\Delta_z=\Delta_e=\Delta_h=\xi_e=\mu=0$, the only non-zero mean-field potentials are $\epsilon^{\rm mf}(\mathbf{k})$ and $\Delta^{\rm mf}(\mathbf{k})$. The main effect of $\epsilon^{\rm mf}(\mathbf{k})$ is the renormalization of $E_G$ to $E_G^R$. Because the densities  are controlled by the gate voltages, we express our results in terms of  $E_G^R$. 
For realistic densities of electrons and holes the system undergoes a second order phase transition as function of temperature, and  below the critical temperature $T_c \sim 0.1 E_0/k_B$ an $s$-wave exciton condensate order parameter $\Delta^{\rm mf}(\mathbf{k})$ appears due to spontaneous symmetry breaking. Because the Hamiltonian has lot of symmetries in the absence of tunneling and spin-orbit couplings, all time-reversal symmetric $s$-wave order parameters $\Delta^{\rm mf}(\mathbf{k})$ with equal total magnitude $\sum_{i=0}^4 |\Delta^{\rm mf}_i (\mathbf{k})|^2$ are degenerate solutions of problem. 

 \begin{figure}[!t]
\includegraphics[width=0.9\linewidth]{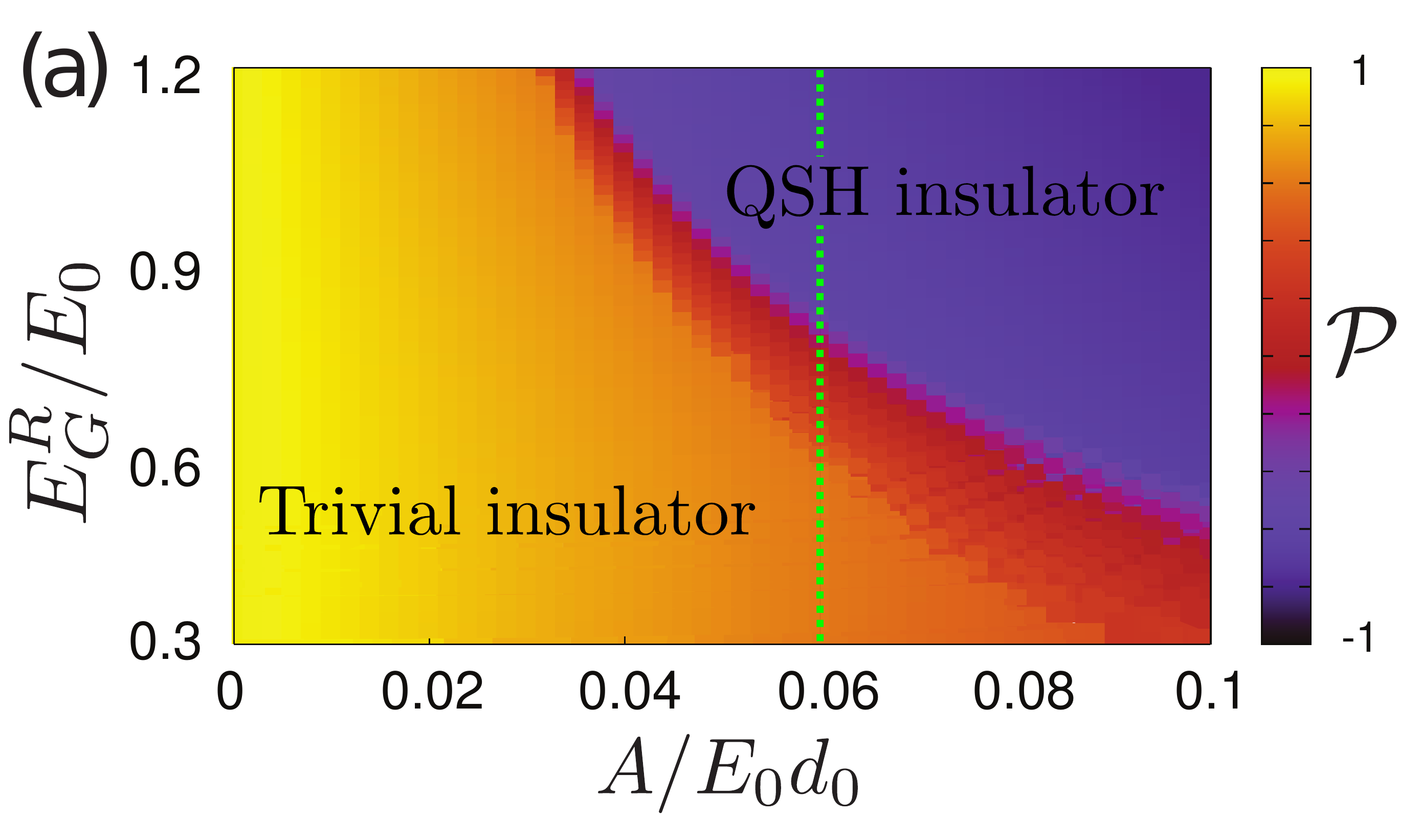}
\includegraphics[width=0.9\linewidth]{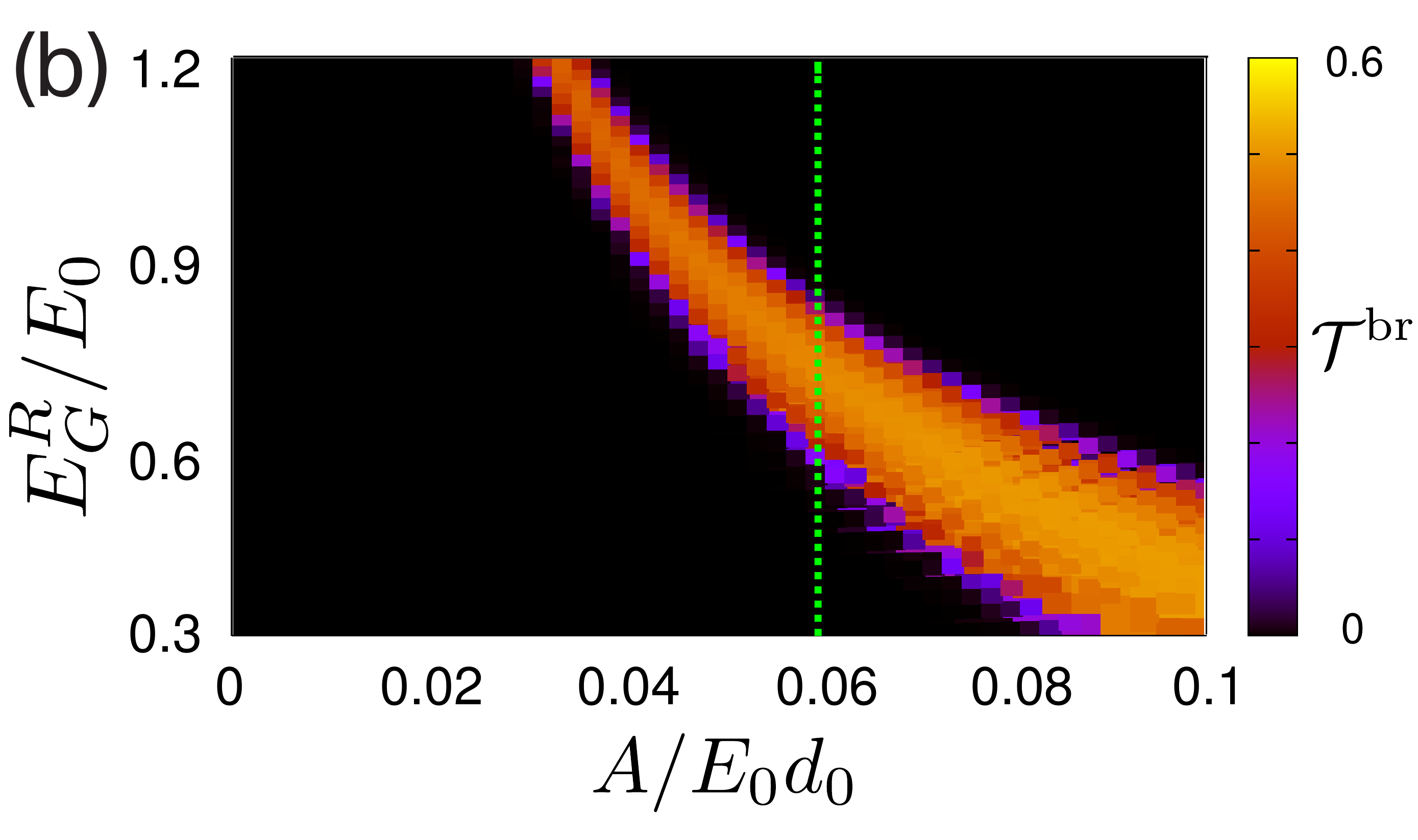}
\includegraphics[width=0.9\linewidth]{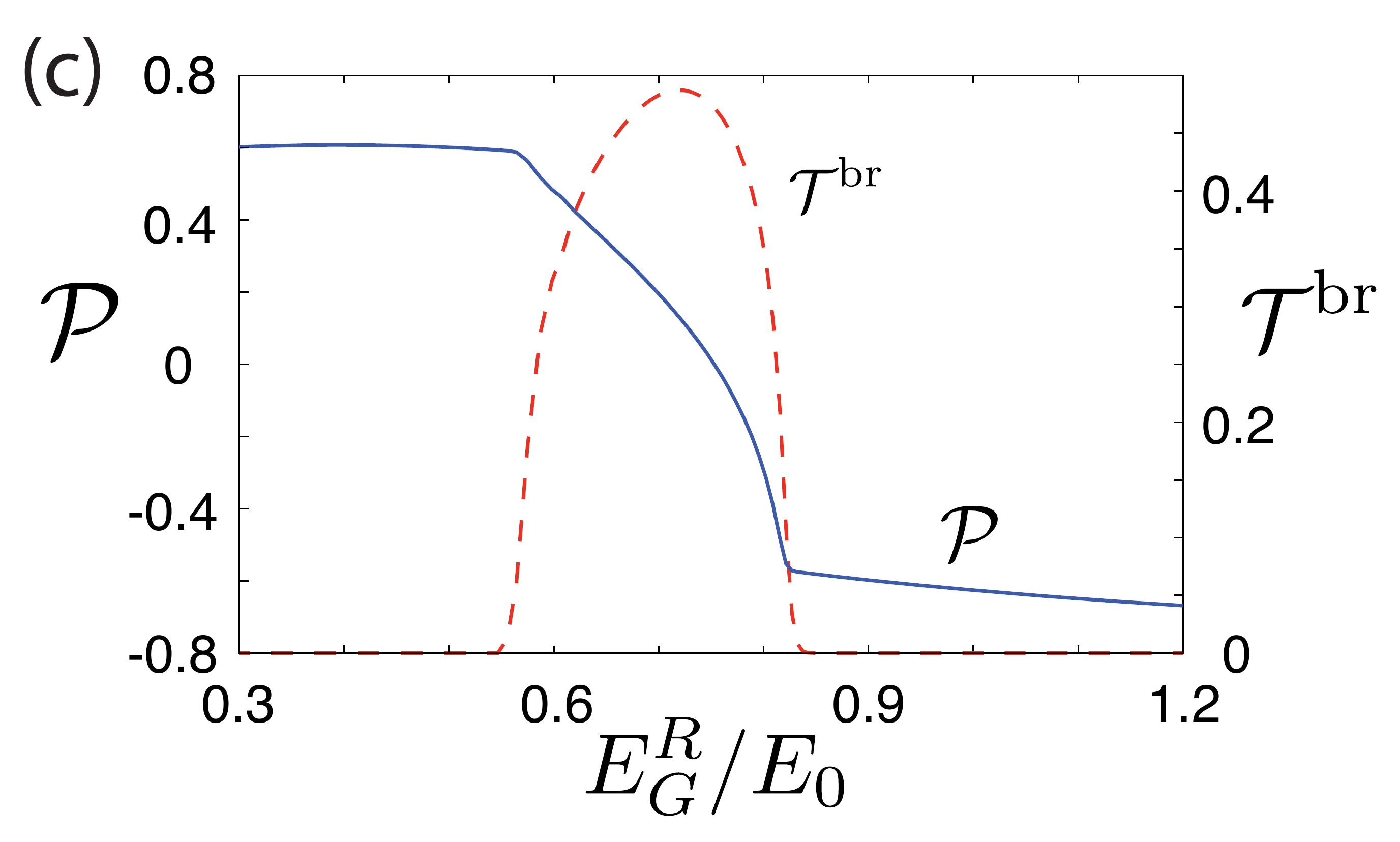}
\caption{(a) Parity of order parameter $\mathcal{P}$ as a function of $E_G^R$ and $A$ for $\Delta_z/E_0=0.02$, $m_e/m_h=1$, $W_{1,2}/d_0=0.8$ and $\mu=d=\Delta_e=\Delta_h=\xi_e=0$. (b) The time-reversal symmetry breaking order parameter $\mathcal{T}^{\rm br}$ as a function of $E_G^R$ and $A$ for the same parameters. (c) Line cut showing $\mathcal{P}$ and $\mathcal{T} ^{\rm br}$ as a function of $E_G^R$ for $A/E_0d_0=0.06$. The insulating phase with broken time-reversal symmetry is separated from the trivial and QSH insulators by second order phase transitions.}
\label{fig:phasediagram}
\end{figure}

Two most important parameters concerning the exciton order parameter are the tunnel couplings $A$ and $\Delta_z$,  because they act as a symmetry breaking terms in the Hamiltonian turning the second order phase transition to the exciton condensate phase into a crossover. It is intuitively clear, that $\Delta_z$ and $A$ favor an even parity exciton condensate order parameter $i\Delta_s \sigma_2$ and an odd-parity exciton order parameter $-\Delta_p(k_x \sigma_3+ik_y\sigma_0)$  ($\Delta_s, \Delta_p \in \mathbb{R}$), respectively.
Therefore, there is a competition between even and odd parity exciton condensate which can be described by studying the parity of order parameter  
\begin{equation}
\mathcal{P}=\frac{\Delta^{\rm even}_{\rm tot}-\Delta^{\rm odd}_{\rm tot}}{\Delta^{\rm even}_{\rm tot}+\Delta^{\rm odd}_{\rm tot}},
\end{equation}
where $\Delta^{\rm even}_{\rm tot}=\sqrt{\int dk \ k  \sum_i |\Delta_{i,0}^{\rm mf}(k)|^2}$ and
$\Delta^{\rm odd}_{\rm tot}=\sqrt{\int dk \  k \sum_{i, n=\pm 1}  |\Delta_{i, n}^{\rm mf}(k)|^2}$ are obtained using expansion  $\Delta_i^{\rm mf}(\mathbf{k})=\sum_n \Delta_{i,n}^{\rm mf}(k) e^{i n \theta_k}$ in terms of  azimuthal angle $\theta_k$ in momentum space.  We obtain the phase diagram in parameters $A$ and $E_G^R$ as they can be changed in a controlled way in the  experiments.  For small $A$ we find a topologically trivial $s$-wave exciton condensate phase, whereas for large $A$ we obtain a topologically non-trivial QSH insulator phase with a $p$-wave exciton order parameter [Fig.~\ref{fig:phasediagram}(a)], in agreement with our expectations. Interestingly, we find that these phases are separated by an insulating phase with spontaneously broken time-reversal symmetry. This time-reversal symmetry broken phase is shown in  Fig.~\ref{fig:phasediagram}(b), where  we have characterized the time-reversal symmetry breaking with a parameter
\begin{equation}
\mathcal{T}^{\rm br}=\frac{\Delta^{\rm br}_{\rm tot}}{\Delta^{\rm ts}_{\rm tot}+\Delta^{\rm br}_{\rm tot}}.
\end{equation}
Here the relative strength of the order parameter obeying the time-reversal symmetry is defined as 
$(\Delta^{\rm ts}_{\rm tot})^2=\int dk \ k\bigg\{ (\Re\Delta_{0,0}^{\rm mf})^2+ \frac{(\Re \Delta_{0,1}^{\rm mf}-\Re \Delta_{0,-1}^{\rm mf})^2}{2}+ \frac{(\Im \Delta_{0,1}^{\rm mf}+\Im \Delta_{0,-1}^{\rm mf})^2}{2} +\sum_{i=1}^3 \bigg[ (\Im\Delta_{i,0}^{\rm mf})^2+  \frac{(\Re \Delta_{i,1}^{\rm mf}+\Re \Delta_{i,-1}^{\rm mf})^2}{2}+   \frac{(\Im \Delta_{i,1}^{\rm mf}-\Im \Delta_{i,-1}^{\rm mf})^2}{2}  \bigg] \bigg\}$ and the strength of time-reversal symmetry-breaking $\Delta^{\rm br}_{\rm tot}$ can be calculated by interchanging the real $\Re \Delta_{i,n}^{\rm mf}$ and imaginary parts $\Im \Delta_{i,n}^{\rm mf}$ of the order parameter in this equation \cite{comment}.
Second order phase transitions are clearly seen at the two boundaries of the time-reversal symmetry broken phase [Fig.~\ref{fig:phasediagram} (c)].  We have numerically confirmed that our results are valid  for chemical potentials $|\mu|/E_0 \lesssim 0.05$.  With increasing $|\mu|$ the difference between the densities of the electrons and holes increases, and the preference for Fermi surface nesting gives rise to  magnetization \cite{Volkov75, Volkov76, MacDonald09}.
Our results are robust against including small spin-orbit coupling $\xi_e/(E_0d_0) = -0.07$ and asymmetry of effective masses $m_e/m_h = 0.84$ that are expected to be present in InAs/GaSb bilayers \cite{Liu-unpublished}, but the locations of the phase boundaries depend strongly on the parameter $\Delta_z$  \cite{Supp}. This situation should be contrasted to HgTe/CdTe QSH insulators described by the same Hamiltonian \cite{Zhang06}, where the exciton condensation does not give rise to phase transitions, because  the conduction and valence bands are localized in the same quantum well so that $A$ is an order of magnitude larger than in InAs/GaSb bilayers \cite{Liu-unpublished}.

Our numerical results can also be interpreted in the light of Ginzburg-Landau theory, which is obtained by  expressing the exciton order parameter as $\Delta^{\rm mf}(\mathbf{k})=i\Delta_s e^{i \phi_s} \sigma_2-\Delta_p e^{i \phi_p} (k_x \sigma_3+ik_y\sigma_0)$ and expanding the free-energy perturbatively using the tunnel couplings and exciton order parameter as a perturbation. We find, that similarly as in Ref.~\cite{Littlewood}, the lowest order terms in the free-energy are proportional to $-\Delta_z \Delta_s \cos \phi_s$ and $-A \Delta_p \cos \phi_p$  favoring the order parameter with $\phi_s=\phi_p=0$. On the other hand, the fourth order expansion contains terms which are proportional to different combinations of $\Delta_z$, $A$, $\Delta_s$ and $\Delta_p$. These terms try to twist the phases of the order parameters $\phi_s$ and $\phi_p$ away from zero, and are thus responsible for the spontaneous breaking of the time-reversal symmetry in the regime of the phase diagram where both $s$- and $p$-wave order parameters are simultaneously large. 

\begin{figure*}[t]
\centerline{\includegraphics[width=0.95\linewidth]{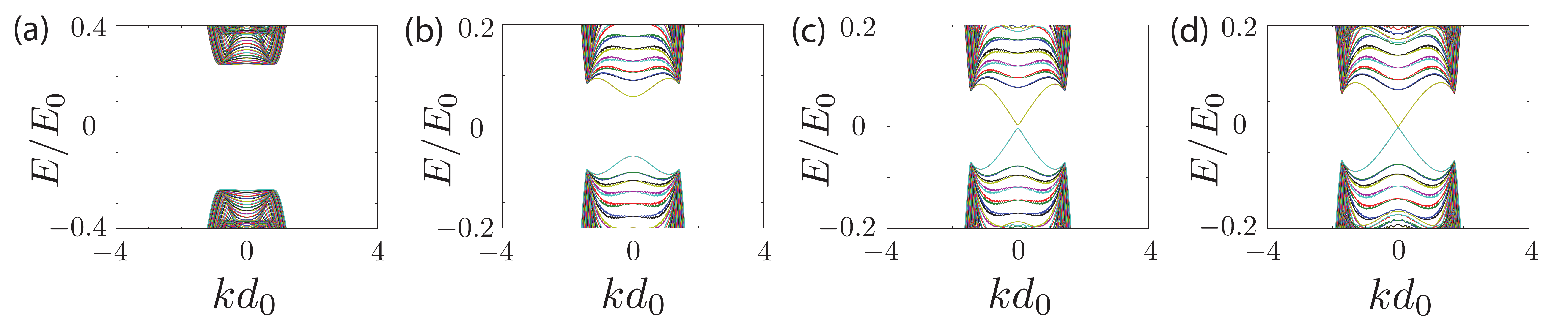}}
\caption{Band structures for (a) $E_G^R=0.3 E_0$, (b) $E_G^R=0.78 E_0$, (c) $E_G^R=0.83E_0$ and $E_G^R=1.11E_0$ and the other parameters same as in Fig.~\ref{fig:phasediagram}(c). Protected helical edge states appear only in the QSH phase.  By decreasing $E_G^R$, the time-reversal symmetry breaking opens a gap in the edge state spectrum, and finally the edge states disappear when one approaches the trivial phase.}
\label{fig_bandstructure}
\end{figure*}

\begin{figure}[tb]
\includegraphics[width=0.88\linewidth]{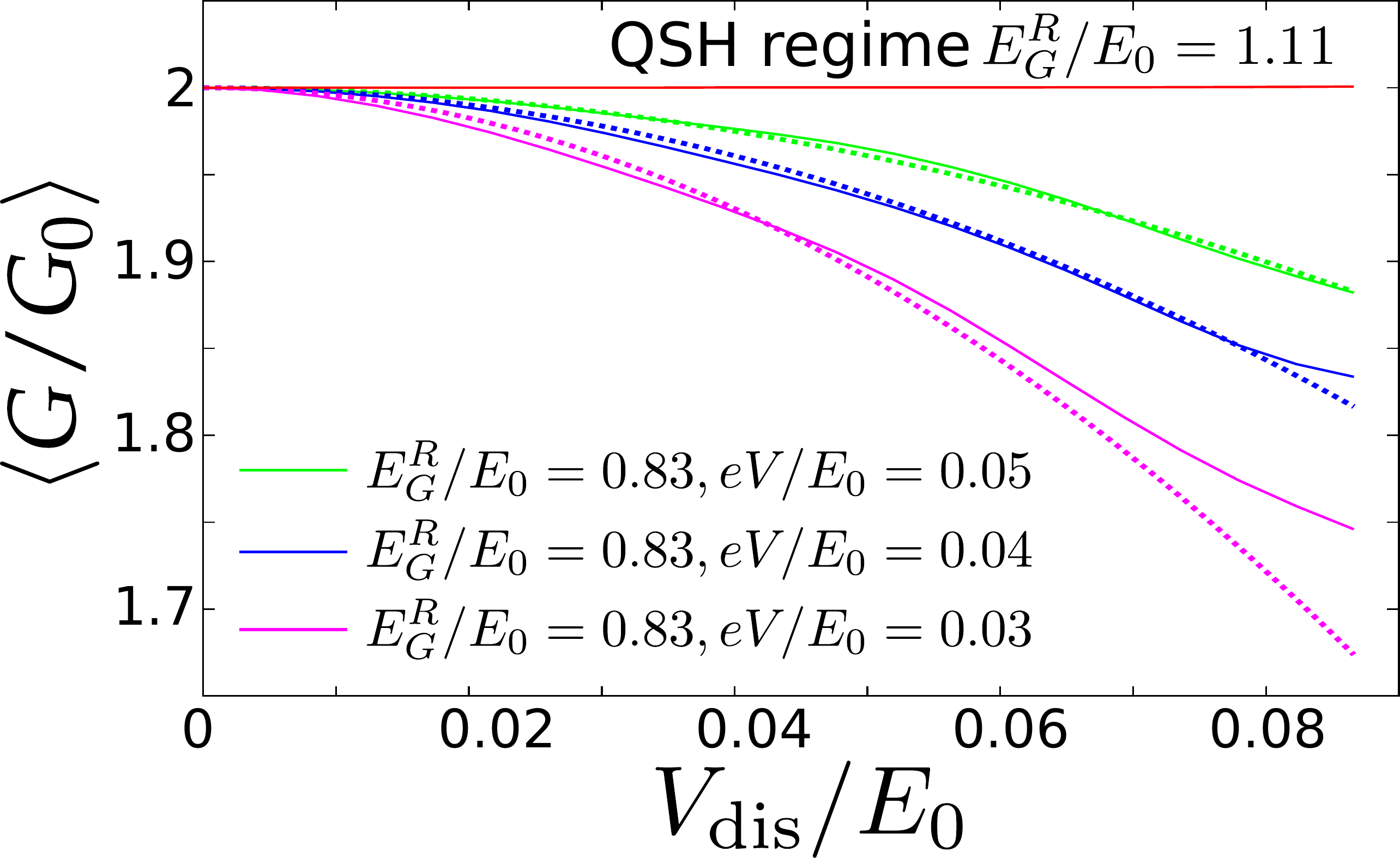}
\caption{Disorder-averaged differential conductance $\langle G \rangle$ as a function of $V_{\rm dis}$ for a device with length $L=100 d_0$ and different values of $E_G^R$ and voltage $eV$. In the QSH regime $E_G^R=1.11 E_0$, the conductance is quantized to $G=2G_0$ ($G_0=e^2/h$). In the regime of weakly-broken time-reversal symmetry $E_G^R=0.83E_0$, numerically calculated $\langle G \rangle$ (thick lines) are in agreement with $G=2G_0 (1-L/\ell)$ (dashed lines), where $\ell$ is obtained from Born approximation with a fitting parameter $a=3.1$. The other parameters same as in Fig.~\ref{fig:phasediagram}(c).}
\label{fig:conductance}
\end{figure}

\textit{Transport.--}To calculate the influence of the exciton condensation on the transport properties of the QSH insulator, we perform a $k\cdot p$ expansion of the mean field potentials, and calculate the conductance using a tight-binding Hamiltonian constructed from the resulting continuum model. The band structures in different parameter regimes are shown in Fig.~\ref{fig_bandstructure} and the results for the disorder-averaged differential conductance $\langle G \rangle$ in Fig.~\ref{fig:conductance}. In the QSH regime, the edge states are protected from elastic back-scattering, and therefore we find perfect conductance quantization for all  disorder strengths $V_{\rm dis}$ shown in the figure. On the other hand, in the regime of weakly-broken time-reversal symmetry the Born approximation gives a mean free path $\ell =a 4\hbar^4v^4k_F^2/(V_{\rm dis}^2 \xi \Delta_{\rm br}^2)$, where we have assumed uncorrelated disorder potential $\langle V(x) V(x') \rangle=V_{\rm dis}^2 \xi \delta(x-x')$ along the edge, $\Delta_{\rm br}$ is the energy gap in the edge state spectrum due to the time-reversal symmetry breaking order parameter, and $a \sim 1$ is a fitting parameter, which depends on the detailed structure of the edge states \cite{edge-structure}. Our numerical results for $\langle G \rangle$  are in good agreement with the Born approximation (Fig.~\ref{fig:conductance}). Importantly, the mechanism of elastic scattering due to the spontaneous breaking of the time-reversal symmetry remains effective for $T \ll T_c$. Since typically $T_c\sim 10$ K, and  we estimate that $\ell \sim 4$ $\mu$m already for a reasonably weak disorder, we conclude that this mechanism is a viable candidate for the explanation of the temperature-independent mean free path observed in recent experiments \cite{Du13, Knez13, Nowack14}. We also predict that the resistance is peaked at the crossing point of the edge state spectrum -- in agreement with the recent experiment \cite{Knez13}, where the maximum resistance was  observed deep inside the topological gap. 

\textit{Summary and discussion.--}In summary, we have studied the exciton condensation in inverted electron-hole bilayers, where the $s$-like conduction band and $p$-like hole band are localized in different quantum well layers. We have calculated the phase diagrams, which show competition between a topologically trivial $s$-wave exciton condensate phase and a non-trivial QSH phase. These topologically distinct phases are separated by an insulating phase with spontaneously broken time-reversal symmetry, which is energetically favoured, because it keeps the system gapped when it experiences a transition between the topologically distinct insulators. Our results can explain the unexpected temperature-independent mean free path observed in InAs/GaSb bilayers \cite{Du13}.

We also point out that a more detailed experimental study  can confirm that the backscattering at the edge happens due to the spontaneous time-reversal symmetry breaking. The phase diagram we discuss can be studied as function of the tunneling amplitudes by controlling the width and height of an insulating barrier between the layers \cite{Murphy94},  and the exciton order parameter can  be probed via the collective modes and vortex excitations. In quantum Hall bilayers,  the exciton order parameter has been studied in terms of counterflow superfluidity and Josephson-like tunneling anomaly \cite{Spielman00, Spielman01, Yoon, Corbino-Eisenstein, Huang12}, and there these properties are known to be strongly influenced by disorder-induced fractionally charged vortices \cite{Moon95, Yang96, Stern01, Balents, Fogler, Hyart, Hyart13, Fertig03, Fertig05, Eastham09, Eastham10, Seradjeh-TIfilms}. A controllable way to open an energy gap in the edge state spectrum by tuning the gate voltage may also be useful for studying Majorana zero modes and for electronic applications.

{\it Acknowledgements.--} The conductance and band structures were calculated using the {\sc kwant}  \cite{kwant}. We have benefited from discussions with I.~Knez and C.~W.~J.~Beenakker. This work was supported by the Foundation for Fundamental Research on
Matter (FOM), the Netherlands Organization for Scientific Research
(NWO/OCW), and the European Research Council.

{\it Note added.} After submission of this work an independent work by Budich {\it et al.} appeared \cite{Trauzettel}, where helical exciton condensates were considered in HgTe bilayers.

\onecolumngrid

\section{Appendix}

\subsection{Description of the order parameters $\mathcal{T}^{\rm br}$ and $\mathcal{P}$}

In this section we explicitly demonstrate that $\mathcal{T}^{\rm br}$ can be used as an order parameter in the description of the spontaneous time-reversal symmetry breaking. Moreover, we show how the parameter $\mathcal{P}$ is related to the $\mathbb{Z}_2$ topological invariant of the two-dimensional time-reversal symmetric insulators. 
Then, we summarize these results in a phase-diagram, which contains the different phases we found in the numerical calculations. 

For the sake of transparency, we use here our numerical observation that the order parameter can be written as
\begin{equation}
\Delta^{\rm mf}(\mathbf{k})=i\Delta_s e^{i \phi_s} \sigma_2-\Delta_p e^{i \phi_p} (k_x \sigma_3+ik_y\sigma_0), \label{ECorderparameter}
\end{equation}  
where $\Delta_s, \Delta_p \in \mathbb{R}$, and we neglect the asymmetry of effective masses $m_e/m_h=1$ and the spin-orbit couplings $\Delta_e$, $\Delta_h$ and $\xi_e$. All these assumptions can be numerically justified, because for realistic values of these parameters the phase-diagram does not change qualitatively. With these assumptions, the mean-field Hamiltonian for $\mu=0$ can be written as 
\begin{eqnarray}
 H&=& \bigg(\frac{\hbar^2 k^2}{2m}-E_G^R \bigg) s_3 \sigma_0 + (\Delta_z+\Delta_s \cos \phi_s) s_2 \sigma_2+(A+\Delta_p \cos\phi_p) \big[s_1 \sigma_3 k_x-s_2 \sigma_0 k_y \big] \nonumber\\ && + \Delta_s \sin\phi_s s_1 \sigma_2+\Delta_p \sin \phi_p  \big[-s_2 \sigma_3 k_x-s_1 \sigma_0 k_y \big], \label{simpH}
\end{eqnarray}
where the pauli matrices $s_i$ and $\sigma_i$ describe the layer and the spin degrees of freedom, respectively. 

The time-reversal symmetry operator is defined as $T=is_0\sigma_2 \cal{K}$, where $\cal{K}$ is the complex conjugation operator. It is easy to see by straightforward calculation that all the terms in the first line of Eq.~(\ref{simpH}) obey the time-reversal symmetry and all the terms in the second line break it. Therefore, all values of phases $\phi_s, \phi_p \ne 0, \pi$ result in spontaneous time-reversal symmetry breaking. On the other hand, using the definition of the order parameter $\mathcal{T}^{\rm br}$ given in the main text and the exciton order parameter given by Eq.~(\ref{ECorderparameter}), we notice that  
\begin{eqnarray}
\mathcal{T}^{\rm br}&=&\frac{\Delta^{\rm br}_{\rm tot}}{\Delta^{\rm ts}_{\rm tot}+\Delta^{\rm br}_{\rm tot}} \nonumber \\
(\Delta^{\rm ts}_{\rm tot})^2&=&\int dk \ k\bigg\{  (\Delta_p k \cos\phi_p)^2  + (\Delta_s \cos \phi_s)^2  \bigg\} \nonumber\\
(\Delta^{\rm br}_{\rm tot})^2&=&\int dk \ k\bigg\{  (\Delta_p k \sin\phi_p)^2 
+ (\Delta_s \sin \phi_s)^2   \bigg\}. \nonumber\\
\end{eqnarray}
Clearly, $\mathcal{T}^{\rm br} \ne 0$ if and only if  $\phi_s, \phi_p \ne 0, \pi$, and therefore $\mathcal{T}^{\rm br}$ can be used as an order parameter for description of the spontaneous time-reversal symmetry breaking. 

\begin{figure}
\includegraphics[width=0.35 \linewidth]{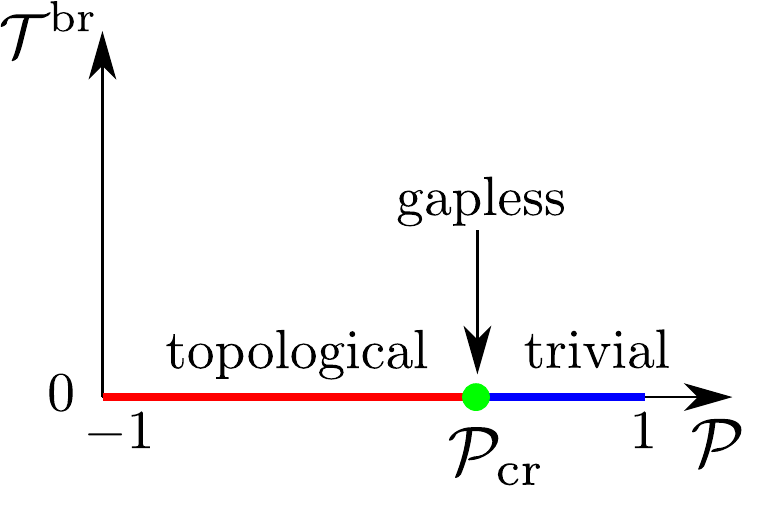}
\caption{Phase-diagram for Hamiltonian (\ref{simpH}) as a function of order parameters $\mathcal{P}$ and $\mathcal{T}^{\rm br}$.  The topological transition between quantum spin Hall and trivial insulators happens at $\mathcal{P}=\mathcal{P}_{\rm cr}$ and $\mathcal{T}^{\rm br}=0$. For $\mathcal{T}^{\rm br}\ne0$ the time-reversal symmetry is spontaneously broken.   For typical parameters $E_G^R=0.7 E_0$, $A=0.06 E_0/d_0$, $\Delta_z=0.02 E_0$ and $\Delta_s+\Delta_p k_F=0.1 E_0$, the critical point is given by $\mathcal{P}_{\rm cr}\approx 0.39$.}
\label{phasediagramPT}
\end{figure}

We now explain the connection between the order parameter $\mathcal{P}$ and $\mathbb{Z}_2$ topological invariant of the two-dimensional time-reversal symmetric insulators. The $\mathbb{Z}_2$ topological invariant is only well-defined in the presence of time-reversal symmetry. Therefore, we set set $\phi_s=\phi_p=0$, so that the Hamiltonian simplfies to a form 
\begin{eqnarray}
 H&=& \bigg(\frac{\hbar^2 k^2}{2m}-E_G^R \bigg) s_3 \sigma_0 + (\Delta_z+\Delta_s) s_2 \sigma_2+(A+\Delta_p ) \big[s_1 \sigma_3 k_x-s_2 \sigma_0 k_y \big]. \label{simpHTRS}
\end{eqnarray}
Furthermore, the order parameter $\mathcal{P}$ defines a relation between $\Delta_s$ and $\Delta_p$ as
\begin{eqnarray}
\mathcal{P}&=&\frac{\Delta^{\rm even}_{\rm tot}-\Delta^{\rm odd}_{\rm tot}}{\Delta^{\rm even}_{\rm tot}+\Delta^{\rm odd}_{\rm tot}}, \nonumber \\ 
(\Delta^{\rm even}_{\rm tot})^2&=&\int dk \ k \ \Delta_s^2 \nonumber \\
(\Delta^{\rm odd}_{\rm tot})^2&=&\int dk \ k \ (\Delta_p k)^2. \label{P-exp}
\end{eqnarray}
These integrals seem to diverge. However, this is only because we have used the low-$k$ expansion of the order parameter instead of taking into account the full momentum dependence of $\Delta_s(k)$ and $\Delta_p(k)$. The relevant scale where the momentum dependent order parameters change must be determined by the parameters $k_F=\sqrt{2m E_G^R}/\hbar$ and  $\hbar/d_0$, which are approximately equal to each other for typical parameters considered in the main text. Therefore, in order to obtain transparent expression for $\mathcal{P}$ we use $\sqrt{2} k_F$ as cut-off momentum in Eqs.~(\ref{P-exp}). This way we obtain
\begin{equation}
\mathcal{P}=\frac{\Delta_s-\Delta_p k_F}{\Delta_s+\Delta_p k_F}.
\end{equation}
In order to describe the connection between the topological invariant, we need to also consider the other parameters $E_G^R$, $A$ and $\Delta_z$ and the overall magnitude of the exciton order parameter $\Delta_s+\Delta_p k_F$. We fix them to the typical values considered in the main text $E_G^R=0.7 E_0$, $A=0.06 E_0/d_0$, $\Delta_z=0.02 E_0$ and $\Delta_s+\Delta_p k_F=0.1 E_0$, and calculate the  $\mathbb{Z}_2$ topological invariant as a function of $\mathcal{P}$ for Hamiltonian (\ref{simpHTRS}). The results are summarized in Fig.~\ref{phasediagramPT}, and can be understood by repeatedly utilizing the powerful result that $\mathbb{Z}_2$ topological invariant can only change when the energy gap of the system closes.

First, we notice that for $\mathcal{P}=1$ the $s$-wave exciton order parameter strongly dominates the tunneling terms and the $p$-wave exciton order parameter. Therefore, in this case it is possible to adiabatically set $A=\Delta_p=0$ without closing the energy gap. By inspecting the Hamiltonian (\ref{simpHTRS}) in this limit, we notice that it is topologically indistinguishable from the usual BCS s-wave superconductor, which is well-known to be topologically trivial i.e.~it does not support edge states. 

Second, in the opposite limit  $\mathcal{P}=-1$ the p-wave exciton order parameter dominates, and it is possible to adiabatically set $\Delta_s=\Delta_z=0$ without closing the energy gap. In this limit the system is described by the Bernevig-Hughes-Zhang Hamiltonian in the topologically nontrivial phase. 

Finally, we find that if we continuously change $\mathcal{P}$ from  $1$  to $-1$ there is exactly one value of $\mathcal{P}$, where the energy gap closes. This means that the $\mathbb{Z}_2$ topological invariant must change at this point, and we have obtained the full phase-diagram shown in Fig.~\ref{phasediagramPT}.

\subsection{Phase diagrams for different $\Delta_z$}

The two most important parameters concerning the exciton order parameter are the tunnel couplings $A$ and $\Delta_z$, because they act as symmetry breaking terms in the Hamiltonian, turning the second order phase transition to the exciton condensate phase into a crossover. As we explained in the main text, these parameters fix the spin structure of the exciton order parameter. However, additionally, they control the phase-boundaries between the different types of exciton condensate phases. 
In the main text, we showed the phase diagram of our model for $\Delta_z/E_0=0.02$, and found three distinct phases (trivial insulator, time-reversal broken insulator and quantum spin Hall insulator phases) in different regimes of the phase-diagram. Here, we address the question how the magnitude of $\Delta_z$ affects the phase-boundaries between these phases. 

\begin{figure}
\includegraphics[width=0.48\linewidth]{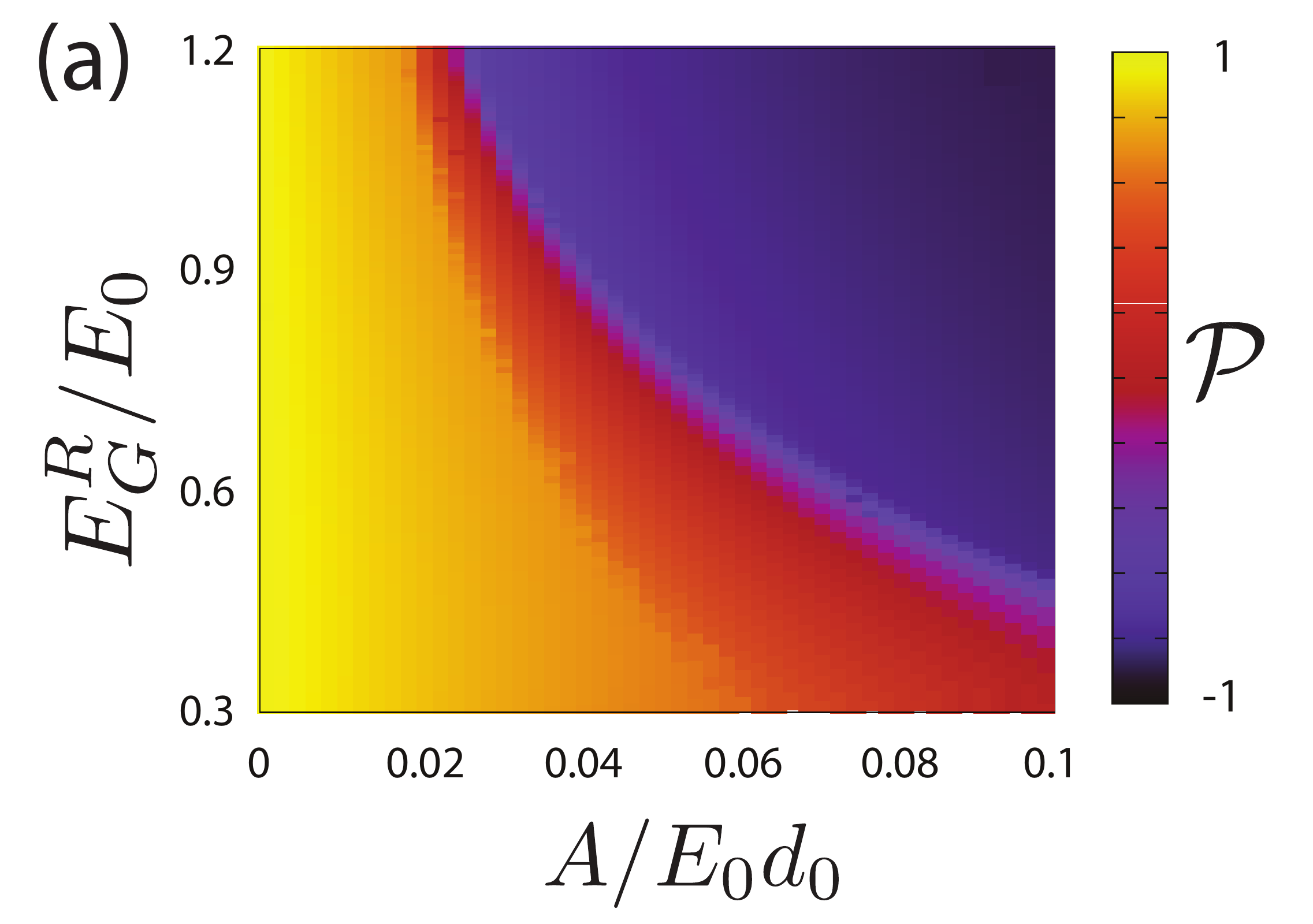}
\includegraphics[width=0.48\linewidth]{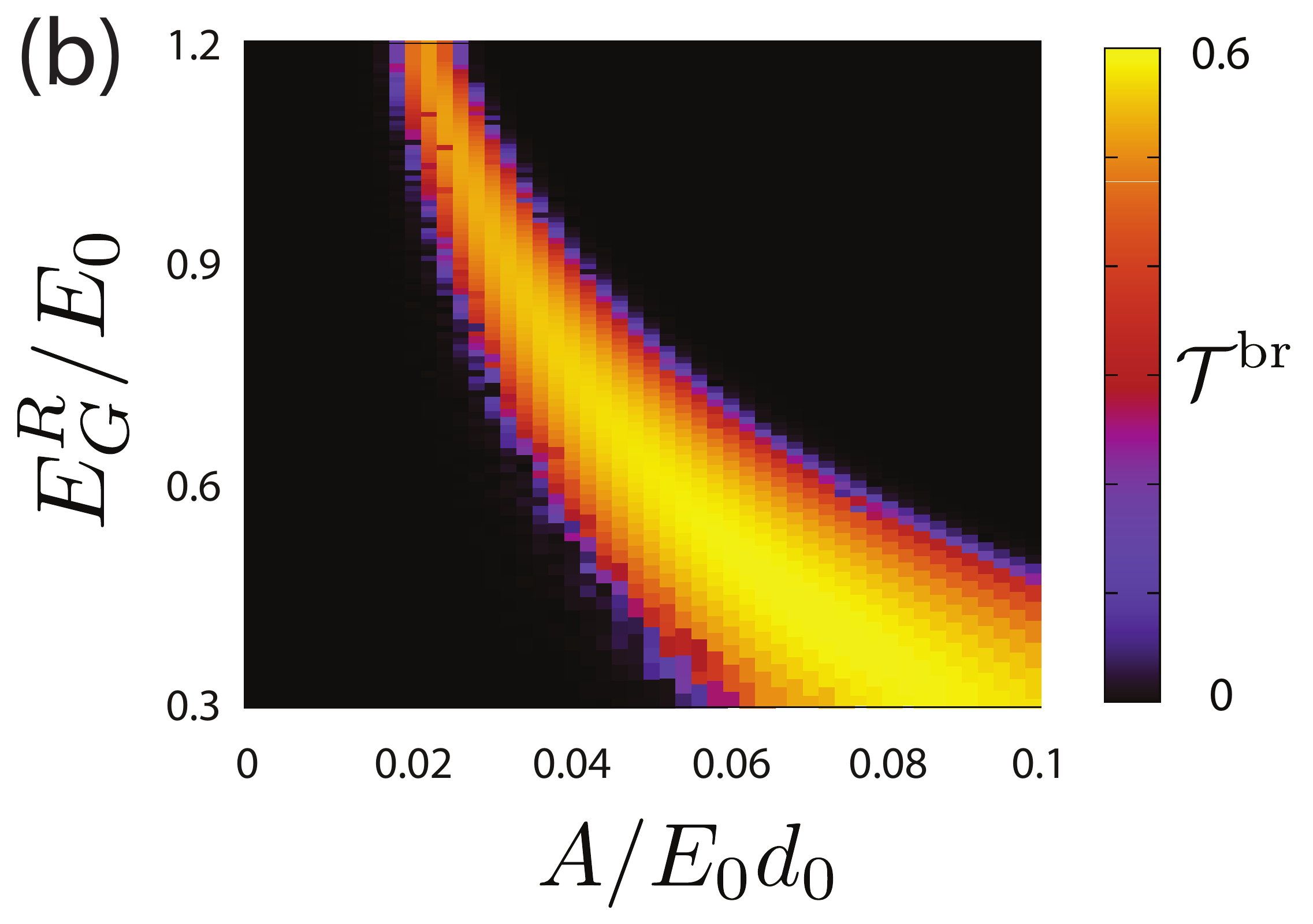}
\includegraphics[width=0.48\linewidth]{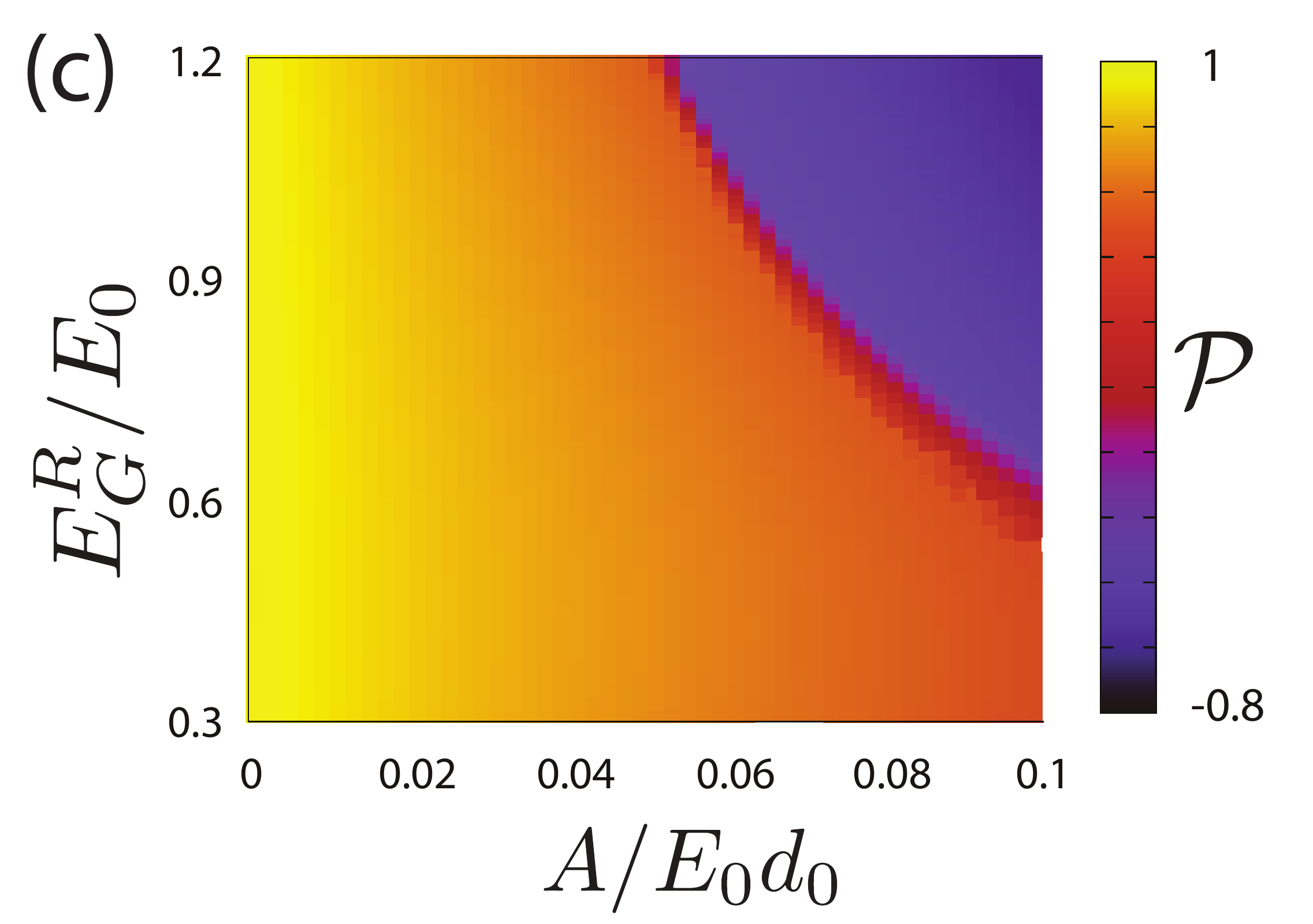}
\includegraphics[width=0.48\linewidth]{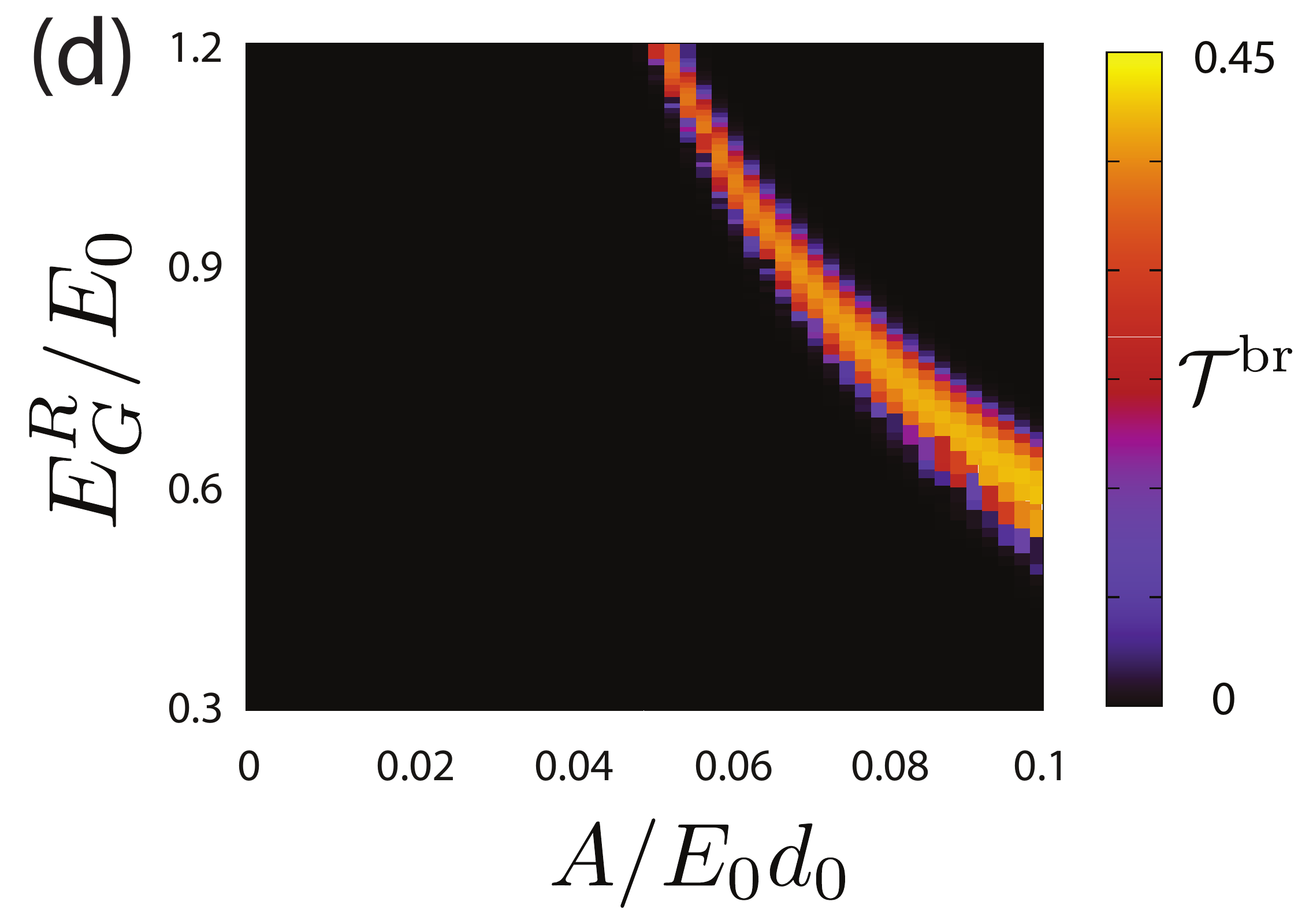}
\caption{Parity of order parameter $\mathcal{P}$  and  the time-reversal symmetry breaking order parameter $\mathcal{T}^{\rm br}$ the as a function of $E_G^R$ and $A$ for  $m_e/m_h=1$, $W_{1,2}/d_0=0.8$ and $\mu=d=\Delta_e=\Delta_h=\xi_e=0$, and different values of $\Delta_z$. (a),(b) $\Delta_z/E_0=0.01$ and (c),(d) $\Delta_z/E_0=0.04$.}
\label{fig:phasediagramsupp}
\end{figure}

The phase-diagrams for different values of $\Delta_z$ are shown in Fig.~\ref{fig:phasediagramsupp}. Based on these figures we conclude that increasing $\Delta_z$ shifts the phase-boundaries to larger values of $A$ as one would have expected. However, this is a reasonably weak effect as long as $\Delta_z \ll k_B T_c \sim  0.1 E_0$, because in this limit the higher order terms in exciton order parameters $\Delta_s$ and $\Delta_p$ in the Ginzburg-Landau theory dominate effect of tunneling. Secondly, the area of the time-reversal broken insulating phase shrinks with increasing $\Delta_z$. We can understand this in the framework of Ginzburg-Landau theory by noticing that the terms $-\Delta_s \Delta_z \cos\phi_s$ and $-A \Delta_p  \cos\phi_p$ try to pin the phases of the s- and p-wave exciton order parameters to zero, and thus for large enough $\Delta_z$ and $A$ it becomes energetically less favorable for the system to spontaneously break the time-reversal symmetry.

\subsection{Temperature dependence of the conductance}

In this section we calculate the temperature dependence of the conductance given by elastic backscattering due to spontaneous time-reversal symmetry breaking. 

As found in the main text the (disorder-averaged) energy-dependent transmission for short samples is given by
\begin{align}
T(E) = (1 - L/\ell(E)),
\end{align} 
where 
\begin{align}
\ell(E) = \frac{4 a \hbar^2 v^2 E^2}{\xi V_{\rm dis}^2 \Delta_{\rm br}^2},
\end{align}
$E$ is the energy with respect to the crossing of the edge state spectrum, and $L$ is the length of the sample. 

Temperature-dependent (disorder-averaged) differential conductance can be obtained from
\begin{align}
G(V, T) = 2 G_0 \int_{-\infty}^{+\infty} dE \frac{1}{4 k_B T} \frac{1}{\cosh^2 \frac{E - eV}{2 k_B T}} T(E) 
\end{align}
Then, by assuming $k_B T \ll eV$, we get
\begin{align}
G(V, T) = 2 G_0 \left(1 - \frac{\xi V_{\rm dis}^2 \Delta_{\rm br}^2 L}{4 a \hbar^2 v^2 e^2 V^2}  - \frac{\pi^2 \xi V_{\rm dis}^2 \Delta_{\rm br}^2 L k_B^2 T^2}{4 a \hbar^2 v^2 e^4 V^4}\right) = G_0 - \Delta G(T=0) - \Delta G(T=0) \left(\frac{\pi k_B T}{e V}\right)^2.
\end{align}

This shows for short samples the temperature independent part of the backscattering dominates at low temperatures, and temperature-dependent corrections can be neglected as long as $k_B T \ll eV$.

In the experiments, the mean free path is measured using long samples. In this case, the Coulomb interactions give rise to an inelastic mean free path and the distribution functions of the electrons  have to be calculated from a solution of a kinetic equation \cite{Bagrets}. We expect that the temperature dependence at small temperatures is further suppressed in long samples, because of the Joule heating.

\end{document}